\documentstyle[aps,twocolumn,amssymb,eqsecnum,tighten,epsf]{revtex}
\begin{document}

\title{Black hole excision with matching}
\author{R. G\'{o}mez, 
	R. L. Marsa and  
	J. Winicour}
\address{
	 Department of Physics and Astronomy,
 	 University of Pittsburgh,
 	 Pittsburgh, PA 15260 }

\date{August 1, 1997}

\maketitle
\widetext
\begin{abstract}

We present a new method for treating the inner Cauchy boundary of a
black hole spacetime by matching to a characteristic evolution. We
discuss the advantages and disadvantages of such a scheme relative to
Cauchy-only approaches. A prototype code, for the spherically symmetric
collapse of a self-gravitating scalar field, shows that matching
performs at least as well as other approaches to handling the inner
boundary.

\end{abstract}


\narrowtext

\section{Introduction}

In many physical systems, boundary conditions are both the most
important and the most difficult part of a theoretical treatment. In
computational approaches, boundaries pose further difficulties. Even
with an analytic form of the correct physical boundary condition in
hand, there are usually many more unstable numerical implementations
than stable ones. Nowhere is the boundary problem more acute than in
the computation of gravitational radiation produced in the coalescence
of two black holes. In order to avoid the topological complications
introduced by the black holes, the proposed strategy for attacking this
problem, initially suggested by W. Unruh \cite{thornburg1987}, is to
excise an interior region surrounded by an apparent horizon. These are
uncharted waters and there are many different tactics that can be
pursued to attain an apparent horizon boundary condition.
\cite{seidelsuen1992,scheel1995a,scheel1995b,marsachoptuik1996,thorn96,cooky90,tod91,nak84,kem91,huq,annin,baum96}
One common feature of all current approaches to this problem
is the use of a Cauchy evolution algorithm in the interior region
bordering the apparent horizon. In this paper we present an alternative
tactic based upon a characteristic evolution in that inner region; and
we present a simple model of its global implementation.

In order to provide orientation, we begin with a synopsis of the
apparent horizon boundary condition and its computational difficulties.
An apparent horizon is the boundary of the region on a Cauchy
hypersurface containing trapped surfaces \cite{wald1984}. This explicit
reference to a Cauchy hypersurface in the definition gives an apparent
horizon an elusive nature. Indeed, there are Cauchy hypersurfaces in
the extended Schwarzschild spacetime which come arbitrarily close to
the final singularity but do not contain an apparent horizon
\cite{waldiyer1991}. There is strong reason to believe that the same is
true in any spherically symmetric black hole spacetime. On the other
hand, when they exist, apparent horizons are useful spacetime markers
because they must lie inside the true event horizon \cite{wald1984}.
Consequently, signals cannot propagate causally from the apparent
horizon to future null infinity ${\cal I}^+$. 

$$~~~~~~~~~~~~$$
\bigskip\bigskip

Thus truncation of the
interior spacetime at the apparent horizon does not affect the
gravitational waves radiated to infinity. This is the physical
rationale behind the apparent horizon boundary condition. 

There is a gauge ambiguity in the inner boundary defined by an apparent
horizon which is associated with the choice of Cauchy foliation. Such
an ambiguity is not associated with the event horizon. However, the
event horizon is of no practical use in a Cauchy evolution since it can
only be constructed in retrospect, after the global geometry of the
spacetime has been determined. A better alternative is the trapping
horizon \cite{sean}, defined as the boundary of the spacetime region
containing trapped surfaces. Here the reference to Cauchy hypersurfaces
is dropped while retaining the quasilocal concept of trapped surfaces.
Trapping horizons exist in any black hole spacetime whereas the
existence of apparent horizons is dependent on the choice of Cauchy
foliation.

In practice, the problem of locating trapped surfaces is partially
solved in the process of setting initial data. For the 3-dimensional
problem of two inspiraling black holes, there are several numerical
approaches for determining appropriate initial Cauchy data
\cite{cooketal93}. An apparent horizon, when it exists, is a marginally
trapped surface and lies on the trapping horizon.  Once the initial
Cauchy hypersurface cuts across a trapping horizon in this way, the
scenario for pathological foliations is not present initially; and a
reasonable choice of lapse should guarantee that future Cauchy
hypersurfaces continue to contain that component of the apparent
horizon. However, in the two black hole problem, besides the two
disjoint apparent horizons present initially, an outer apparent horizon
(surrounding them) is expected to form at a later time. Finding and
locating this outer apparent horizon can make the computational problem
enormously easier by using it as the new inner boundary at this stage.
Excellent progress has been made in designing apparent horizon finders
and trackers for this purpose. However, it is not known what lapse
condition on a Cauchy foliation would guarantee that an outer apparent
horizon form at the earliest possible time.

Besides these geometrical issues there are a number of serious
computational difficulties in implementing an apparent horizon boundary
condition. In order to obtain gravitational waveforms, the
computational domain must cover a time interval of the order of several
hundred $M$ in the exterior region whereas typically a singularity
forms on a time of order $M$ in the region close to the apparent
horizon. Thus a slicing which avoids the singularity for several
hundred $M$ will necessarily develop coordinate singularities.  In
addition, the inner boundary traced out by an apparent horizon is
generically spacelike (at best lightlike). Thus if the coordinates
defining the numerical grid were to remain constant in time on the
boundary (``apparent horizon locking'') then the coordinate
trajectories would have to be superluminal. While horizon locking works
in the spherically symmetric case
\cite{seidelsuen1992,scheel1995a,scheel1995b,marsachoptuik1996}, it is
difficult to implement in a Cartesian 3-dimensional grid. The
alternative is to let the apparent horizon move through the coordinate
grid.  At the same time, the location of the apparent horizon must be
determined by solving an elliptic equation or an equivalent extremum
problem.  The requirements on the grid are further complicated when the
black hole is spinning. On top of all these difficulties, the
computational techniques must ensure that the strong fields inside the
apparent horizon boundary do not severely leak into the exterior due to
finite difference approximations. Causal differencing
\cite{seidelsuen1992} and algorithms based upon a strictly hyperbolic
version of the initial value problem \cite{abrah95} have been proposed
to avoid this. However, no 3-dimensional Cauchy code has yet been
successful in evolving a Schwarzschild black hole.

\begin{figure}
\centerline{\epsfxsize=3.2in\epsfbox{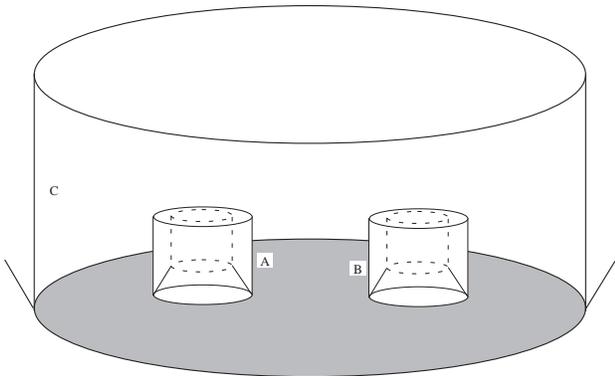}}
\caption{
A schematic of how matching to ingoing null cones could be used 
with two black holes.  The inner Cauchy evolution is matched at an 
outer world tube $C$ to a null evolution on outgoing null cones; and 
at two interior world tubes $A$ and $B$, to null evolutions on ingoing 
null cones.  The evolutions on the ingoing null cones stop at the 
apparent horizons (dotted lines) which surround the two black holes.
}
\label{fig:tblackholes}
\end{figure}

It is clear that the 3-dimensional coalescence of black holes
challenges the limits of computational know-how. We wish to present
here a new approach for excising an interior trapped region which might
provide enhanced flexibility in tackling this important problem. In
this approach, we locate the interior boundary of the Cauchy evolution
{\em outside} the apparent horizon. Across this inner Cauchy boundary
we match to a characteristic evolution based upon an ingoing family of
null hypersurfaces. It is the inner boundary condition for the
characteristic evolution which is then given by a null hypersurface
version of the apparent horizon boundary condition. In the case of two
black holes, the inner boundary would consist of two disjoint
topological spheres, chosen so that their inner directed null normals
are converging. FIG \ref{fig:tblackholes} provides a schematic picture 
of the global
strategy. Two disjoint characteristic evolutions, based upon ingoing
null hypersurfaces, are matched across worldtubes A and B to a Cauchy
evolution of the shaded region. The interior boundary of each of these
characteristic evolutions borders a region containing trapped surfaces.
The outer boundary of the Cauchy region is another worldtube C, which
matches to an exterior characteristic evolution based upon outgoing
null hypersurfaces extending to null infinity.

This strategy offers several advantages in addition to the possibility
of restricting the Cauchy evolution to the region outside the black
holes.  Although, finding a marginally trapped
surface on the ingoing null hypersurfaces remains an elliptic problem,
there is a natural radial coordinate system $(r,\theta,\phi)$ to
facilitate its solution.  However it is also possible to locate a
trapped surface on the ingoing null hypersurface by a purely algebraic
condition. Since this trapped surface (when it exists) lies in the region
invisible to ${\cal I}^+$ it can be used to replace the trapping
horizon as the inner boundary. In either case, moving the black hole
through the grid reduces to a 1-dimensional radial motion, leaving the
angular grid intact and thus reducing the complexity of the
computational masks which excise the inner region. (The angular
coordinates can even rotate relative to the Cauchy coordinates in order
to accommodate spinning black holes). The chief problem of this
approach is that a caustic may be encountered on the ingoing null
hypersurface before entering the trapped region. This is again a
problem whose solution lies in choosing the right initial data and also
the right geometric shape of the two-surface across which the Cauchy
and characteristic evolutions are matched. There is a great deal of
flexibility here because of the important feature that initial data can
be posed on a null hypersurface without constraints.

The strategy of matching an interior Cauchy evolution to an exterior
{\em outgoing} characteristic evolution has been described
\cite{And,Bis,bis2} and implemented to provide a computational Cauchy
outer boundary condition in various cases, ranging from 1 and 2
dimensional simulations \cite{Clarke,Cla95,Dub95,gomez1996}
to 3-dimensional simulations that include ${\cal I}^+$
\cite{bishop1996,jcp}.  A slight modification allows changing an
outgoing null formalism (and its evolution code) to an ingoing one.
This is briefly reviewed in Sec.~\ref{sec:null}. By matching Cauchy and
characteristic algorithms at both an inner and outer boundary, the
ability to include ${\cal I}^+$ facilitates locating the true event
horizon while excising an interior trapped region. In
Sec.~\ref{sec:trapped}, we discuss the problem of locating trapped
surfaces on an ingoing null hypersurface. In Sec.~\ref{sec:spherical},
we present an implementation of these ideas to the global evolution of
spherically symmetric, self gravitating scalar waves propagating in a
black hole spacetime. In this case, the performance of the matching
approach equals that of previous Cauchy-only schemes that have been
applied to this problem \cite{scheel1995a,scheel1995b,marsachoptuik1996,anninos1995}.

\section{Cauchy-characteristic matching} \label{sec:null}

\subsection{The null formalism}

We introduce a unified formalism for coordinates based upon either
ingoing or outgoing null hypersurfaces. Let $w$ label these
hypersurfaces, $x^A$ ($A=2,3$), be labels for the null rays and $r$ be
a surface area distance. In the resulting $x^\alpha=(w,r,x^A)$
coordinates, the metric has the Bondi-Sachs
form~\cite{vanderburg1962,sachs}
\begin{eqnarray}
   ds^2 &=& g_{ww} dw^2
        +2g_{wr}dwdr +2g_{wA}dwdx^A \nonumber \\ && +g_{AB}dx^Adx^B,
   \label{eq:wmet}
\end{eqnarray}
where $det(g_{AB})=r^2 det(q_{AB})=r^2 q$, with $q_{AB}$ a unit sphere metric.
In the outgoing case, writing $w=u$, it is convenient to express the metric
variables in the form 
\begin{eqnarray}
   ds^2=&-&\left(e^{2\beta}{V \over r} -r^2h_{AB}U^AU^B\right)du^2
        -2e^{2\beta}dudr 
			\nonumber \\ &-& 2r^2 h_{AB}U^Bdudx^A +r^2h_{AB}dx^Adx^B,
   \label{eq:umet}
\end{eqnarray}
where $h^{AB}h_{BC}=\delta^A_C$. This yields
the standard outgoing null coordinate version of the Minkowski metric
by setting $\beta=U^A=V-r=h_{AB}-q_{AB}=0$. In the ingoing
case, writing $w=v$, the only component of the Minkowski metric which
differs is $g_{vr}=-g_{ur}$. This can be effected by the substitution
\begin{equation}
      \beta\rightarrow \beta+i\pi/2 
     \label{eq:betasub}
\end{equation}
in the outgoing form of the metric.

The substitution (\ref{eq:betasub}) can also be used in the curved
space case to switch from outgoing to ingoing coordinates, in which
case it is equivalent to an imaginary shift in the integration constant
for the Einstein equation determining $\beta$ (see equation
(\ref{eq:beta}) below).  This leads to the ingoing version of the
metric 
\begin{eqnarray}
   ds^2 &=& \left(e^{2\beta}{V \over r} +r^2h_{AB}U^AU^B\right)dv^2
	+2e^{2\beta}dvdr \nonumber \\ && -2r^2 h_{AB}U^Bdvdx^A +r^2h_{AB}dx^Adx^B.
   \label{eq:vmet} 
\end{eqnarray} Of course, at a given space-time point the values of the
coordinates $r$ and $x^A$ and the metric quantities $\beta$, $U^A$, $V$ and
$h_{AB}$ are not the same in the ingoing and outgoing cases; but since we
do
not consider transformations between outgoing and ingoing coordinates there
is no need to introduce any special notation to distinguish between them.

This same substitution also provides a simple switch from the outgoing
to the ingoing version of Einstein equations  $G_{\alpha \beta}=8\pi
T_{\alpha \beta}$ written in null coordinates. This is consistent
because $\beta$ contains a free integration constant which can be
chosen to be complex (as long as it leads to a real metric). In order
to see how this works consider the outgoing version of the null
hypersurface equations~\cite{winicour1983,winicour1984}:
\begin{eqnarray}
\beta_{,r} &=& \frac{1}{16}rh^{AC}h^{BD}h_{AB,r}h_{CD,r}
         \nonumber \\ && +2\pi r T_{rr},
\label{eq:beta}
\\
(r^4e^{-2\beta}h_{AB}U^B_{,r})_{,r}  &=&
2r^4 \left(r^{-2}\beta_{,A}\right)_{,r}
- r^2 h^{BC}D_{C}h_{AB,r} \nonumber \\ && +16\pi r^2 T_{rA}
\label{eq:u}
\\
2e^{-2\beta}V_{,r} &=& {\cal R} - 2 D^{A} D_{A} \beta
-2 D^{A}\beta D_{A}\beta 
\nonumber \\ && +r^{-2} e^{-2\beta} D_{A}(r^4U^A)_{,r}
\nonumber \\ && -\frac{1}{2}r^4e^{-4\beta}h_{AB}U^A_{,r}U^B_{,r} 
\nonumber \\ && +8\pi r^2 (T-g^{AB}T_{AB}),
\label{eq:v}
\end{eqnarray}
where $D_A$ is the covariant derivative and ${\cal R}$ the curvature
scalar of the 2-metric $h_{AB}$.  The $\beta$-equation (\ref{eq:beta})
allows the substitution (\ref{eq:betasub}) to be regarded as a change
in integration constant. Then carrying out this substitution in equations
(\ref{eq:beta})-(\ref{eq:v}) leads to the ingoing version of the null
hypersurface equations:
\begin{eqnarray}
\beta_{,r} &=& \frac{1}{16}rh^{AC}h^{BD}h_{AB,r}h_{CD,r}
         \nonumber \\ && +2\pi r T_{rr},
\label{eq:ibeta}
\\
-(r^4e^{-2\beta}h_{AB}U^B_{,r})_{,r}  &=&
2r^4 \left(r^{-2}\beta_{,A}\right)_{,r}
-r^2 h^{BC}D_{C}h_{AB,r} \nonumber \\ && + 16\pi r^2 T_{rA}
\label{eq:iu}
\\
-2e^{-2\beta}V_{,r} &=& {\cal R} - 2 D^{A} D_{A} \beta
-2 D^{A}\beta D_{A}\beta 
\nonumber \\ && -r^{-2} e^{-2\beta} D_{A}(r^4U^A)_{,r}
\nonumber \\ && -\frac{1}{2}r^4e^{-4\beta}h_{AB}U^A_{,r}U^B_{,r} 
\nonumber \\ && +8\pi r^2 (T-g^{AB}T_{AB}).
\label{eq:iv}
\end{eqnarray}
This formal substitution also applies to the dynamical equations
and provides a simple means to switch between evolution algorithms
based upon ingoing and outgoing null cones.

As we have already noted, although the same coordinate labels $r$ and
$x^A$ are used for notational simplicity in  both the outgoing metric
(\ref{eq:umet}) and the ingoing metric (\ref{eq:vmet}), they represent
different fields. An exception occurs for spherical symmetry where the
the surface area coordinate $r$
can be defined uniquely in terms of the same 2-spheres of
symmetry used in both the ingoing and the outgoing coordinates. In
this case, the spacelike or timelike character of the $r=const$
hypersurfaces is consistent under the substitution (\ref{eq:betasub})
because the change involved in going from (\ref{eq:v}) to (\ref{eq:iv})
implies that $V$ changes sign in switching from outgoing to ingoing
coordinates. As a result we obtain a consistent value for
$g^{\alpha\beta}r_{,\alpha}r_{,\beta}=\pm e^{-2\beta}V/r$, with the $+$
($-$) sign holding for outgoing (ingoing) coordinates.

In the absence of spherical symmetry, the surface area coordinate $r$
used in the Bondi-Sachs formalism has a gauge ambiguity associated with
the changes in ray labels $x^A\rightarrow y^A(x^B)$, under which it
transforms as a scalar density. On any null hypersurface with a
preferred compact spacelike slice ${\cal S}_0$, this coordinate freedom
in $r$ may be fixed by requiring that $r=const$ on ${\cal S}_0$. This
then determines a unique $r=const$ foliation on either the ingoing or
outgoing null hypersurface emanating from ${\cal S}_0$.

\subsection{Matching}

Cauchy-characteristic matching can be used to replace artificial boundary
conditions which are otherwise necessary at the outer boundary of a
finite Cauchy domain. The exterior characteristic evolution can then be
extended to null infinity to form a globally well-posed initial value
problem. In tests of nonlinear 3-dimensional scalar waves,
Cauchy-characteristic matching dramatically outperforms the best
available artificial boundary condition both in accuracy and
computational efficiency \cite{bishop1996,jcp}.

We now describe how this matching strategy can be used at the inner
boundary of a Cauchy evolution which is joined to an ingoing null
evolution. On the initial Cauchy hypersurface, denoted by time $t_0$,
let ${\cal S}_0$ be a (topological) 2-sphere forming the inner boundary
of the region being evolved by Cauchy evolution. Let ${\cal W}$
represent the future evolution of ${\cal S}_0$ under the flow of the
vector field $t^a = \alpha n^a +\beta ^a$, where $n^a$ is the unit
vector field normal to the Cauchy hypersurfaces $t=const$ and $\alpha$
and $\beta^a$ are the lapse and shift. Given the initial Cauchy data on
$t_0$, boundary data must be given on the world tube ${\cal W}$ in
order to determine its future evolution. The Cauchy hypersurfaces
foliate this world tube into spheres ${\cal S}_t$. The boundary data is
obtained by matching across ${\cal W}$ to an interior null evolution
based upon the null hypersurfaces $v=const$ emanating inward from from
the spheres ${\cal S}_t$. The evolutions are synchronized by setting
$v=t$ on ${\cal W}$.

\begin{figure}
\centerline{\epsfxsize=3.2in\epsfbox{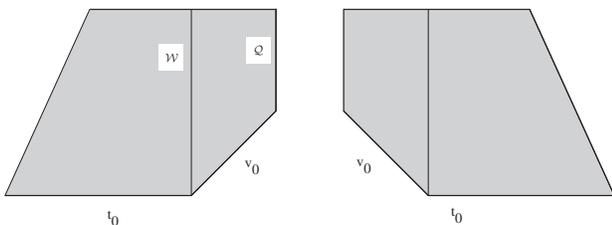}}
\caption{
The shaded region is the domain of dependence of initial data which is given on
a Cauchy slice $t_0$ and an ingoing null slice $v_0$.  $\cal Q$ is the 
apparent horizon.  $\cal W$ is the matching world tube.
}
\label{fig:domain}
\end{figure}

The combination of initial null data on $v_0$ and initial Cauchy data on
$t_0$ determine the future evolution in their combined domain of
dependence as illustrated in FIG \ref{fig:domain}. In order to avoid 
dealing with
caustics we terminate the null hypersurfaces $v_t$ on an inner boundary
${\cal Q}$ whose location is determined by a trapping condition, as
discussed below. This inner boundary plays a role analogous to an
apparent horizon inner boundary in a pure Cauchy evolution. The region
inside ${\cal Q}$ is causally disjoint from the domain of dependence
which is evolved from the initial data. In the implementation of this
strategy in the model spherically symmetric problem of Sec.
\ref{sec:spherical}, we describe in detail how data is passed back and
forth across ${\cal W}$ to supply an outer boundary value for the null
evolution and an inner boundary value for the Cauchy evolution. In the
remainder of this section, we discuss how some of the key underlying
issues might be handled in the absence of symmetry.

In order to ensure that an inner trapping boundary exists it is
necessary to choose initial data which guarantees black hole
formation.  Such data can be obtained from initial Cauchy data on $t_0$
for a black hole. However, rather than extending the Cauchy
hypersurface inward to the apparent horizon it could instead be
truncated at an initial matching surface ${\cal S}_0$ located
sufficiently far outside the apparent horizon to avoid computational
problems with the Cauchy evolution. The initial Cauchy data would then
be extended into the interior of ${\cal S}_0$ as null data on $v_0$
until a trapping boundary is reached. Two ingredients are essential in
order to arrange this.  First, ${\cal S}_0$ must be chosen to be
convex, in the sense that its outward null normals uniformly diverge
and its inner null normals uniformly converge. Given any physically
reasonable matter source, the focusing theorem then guarantees that the
null hypersurface $v_0$ emanating inward from ${\cal S}_0$ continues to
converge until reaching a caustic. Second, initial null data must be
found which leads to trapped surfaces on $v_0$ before such a caustic is
encountered. The existence of trapped surfaces depends upon the
divergence of the outward normals to slices of $v_0$. Given the
appropriate choice of ${\cal S}_0$ the existence of such null data is
guaranteed by the evolution of the extended Cauchy problem. However, it
is not necessary to actually carry out such a Cauchy evolution to
determine this null data. It is the data on ${\cal S}_0$ which is most
critical in determining whether a black hole can form. This can be
phrased in terms of the trapping gravity of ${\cal S}_0$, introduced
below. In the spherically symmetric Einstein-Klein-Gordon model (see
Sec. \ref{sec:spherical}), if ${\cal S}_0$ has sufficient trapping
gravity to form a trapped surface on $v_0$ in the absence of scalar
waves crossing $v_0$ then a trapped surface also forms in the presence
of scalar waves. In the absence of symmetry, this suggests that given
appropriate initial Cauchy data for horizon formation that the simplest
and perhaps physically most relevant initial null data for a black hole
would correspond to no gravitational waves crossing $v_0$. The initial
null data on $v_0$ can be posed freely, i.e. it is not subject to any
elliptic or algebraic constraints other than continuity requirements
with the Cauchy data at ${\cal S}_0$. In the vacuum case, the absence
of gravitational waves in the null data has a natural (although
approximate) formulation in terms of setting the ingoing null component
of the Weyl tensor to zero on $v_0$.

Key to the success of this approach is the proper trapping behavior,
i.e.  the convergence of both sets of null vectors normal to a set of
slices of $v_0$ located between the caustics and the matching boundary.
By construction, the ingoing null hypersurface ${\cal N}_0$, given by
$v=v_0$, is converging along all rays $x^A$ leaving the initial slice
${\cal S}_0$ coordinatized by $r=R_0$. (Here $(v,r,x^A)$ are ingoing
null coordinates.) In order to investigate the trapping of ${\cal N}_0$
we must determine the divergence of slices ${\cal S}$ of ${\cal N}_0$
defined by $r=R(x^A)$. Let $n^{\alpha}$ be tangent to the generators of
${\cal N}_0$, with normalization $n_{\alpha}=-g_{vr} v_{,\alpha}$. Then
$n^{\alpha}v_{,\alpha}=n^{\alpha }x^A_{,\alpha}=0$ and
$n^{\alpha}r_{,\alpha}=-1$. Let $l_{\alpha}$ be the outgoing normal to
$S$, normalized by $n^{\alpha}l_{\alpha}=-1$. Then
\begin{equation} l_{\alpha} =
-l v_{,\alpha} +r_{,\alpha} - R_{,A}x^A_{,\alpha}
\end{equation}
where
\begin{equation}
l = {1\over 2}g_{vr}(g^{rr}-2g^{rA}R_{,A} +
g^{AB}R_{,A}R_{,B}).
\end{equation}
Its contravariant components are
\begin{equation}
l^v = g^{vr}
\end{equation}
\begin{equation}
l^r = {1\over
2}g^{rr}-{1\over 2}g^{AB}R_{,A}R_{,B} \end{equation} and \begin{equation}
l^A
= g^{rA} -g^{AB}R_{,B}.
\end{equation}

Let $\gamma^{\alpha}_{\beta}=g^{\alpha}_{\beta}+n^{\alpha}l_{\beta}
+l^{\alpha}n_{\beta}$ be the projection tensor into the tangent space
of ${\cal S}$ and define $\gamma^{\alpha\beta}$  and
$\gamma_{\alpha\beta}$ by raising and lowering indices with
$g_{\alpha\beta}$. Its contravariant components are $\gamma^{\alpha
v}=0$, $\gamma^{AB}=g^{AB}$, $\gamma^{rA}=g^{AB}R_{,B}$ and
$\gamma^{rr}=g^{AB}R_{,A}R_{,B}$; and its covariant components are
$\gamma_{\alpha r}=0$, $\gamma_{AB}=g_{AB}$, $\gamma_{vA}=
g_{vr}R_{,A}+g_{vA}$ and
$\gamma_{vv}=g^{AB}(g_{vr}R_{,A}+g_{vA})(g_{vr}R_{,B}+g_{vB})$.

The outward divergence of ${\cal S}$ is given by
$\Theta_l=2\gamma^{\alpha\beta}\nabla_{\alpha}l_{\beta}$. (The conventions
are chosen so that $\Theta=2/r$ for a $r=const$ slice of an outgoing null
cone in
Minkowski space.) Then a straightforward calculation yields
\begin{eqnarray}
   {1\over 2}\Theta_l &=&  {1\over r}g^{rr}
                -{g^{vr}\over \sqrt{q}}[\sqrt{q} g^{AB}(g_{vr}R_{,B}+g_{vB})]_{,A}
   \nonumber \\ && -{g^{vr}\over r}(rg_{vr}g^{AB})_{,r}R_{,A}R_{,B}
   \nonumber \\ && -g^{vr}R_{,B}(g^{AB}g_{vA})_{,r},
    \label{eq:divergence}
\end{eqnarray}
which is to be evaluated on ${\cal S}$ after the derivatives are
taken.  The divergence of the generators tangent to  ${\cal N}_0$ is
given by $\Theta_n=\gamma^{\alpha\beta}\nabla_{\alpha}n_{\beta}$.
Then $\Theta_n=-2/r$ so that ${\cal N}_0$ is converging in accord
with our construction.

\section{Trapping}  \label{sec:trapped}

A slice of ${\cal N}_0$ is trapped if $\Theta_l<0$. In terms of
the ingoing Bondi metric variables defined in (\ref{eq:vmet}),
equation (\ref{eq:divergence}) gives
\begin{eqnarray}
  {r^2 e^{2\beta}\over 2}\Theta_l = &-&V
       -{1\over \sqrt{q}}[\sqrt{q}(e^{2\beta}h^{AB}R_{,B}-r^2 U^A)]_{,A}
 \nonumber \\ &-& r(r^{-1}e^{2\beta}h^{AB})_{,r}R_{,A}R_{,B}
                +r^2R_{,A}U^A_{,r}.
    \label{eq:diverg}
\end{eqnarray} 
Setting $\Theta_l=0$ in equation (\ref{eq:diverg}) gives a
2-dimensional Laplace equation for the function $R(x^A)$ which locates
a marginally trapped surface ${\cal A}$. Such a surface lies on a
trapping horizon and is (a component of) the apparent horizon of any
Cauchy hypersurface which contains it.

For a marginal surface to lie on an {\em outer} trapping horizon its
trapping gravity, defined as \cite{sean}
\begin{equation}
    \kappa= \sqrt{-{1\over 8}{\cal L}_n \Theta_l},
   \label{eq:gravity}
\end{equation}
must be real and positive, so that $\kappa \ge 0$. The trapping gravity
generalizes the concept of the surface gravity of an event horizon to
trapping horizons. In our coordinate system,
$n^{\alpha}\partial_{\alpha}=-\partial_r$. Thus if the surface
$r=R(x^A)$ is marginally trapped then positive trapping gravity implies
that the surface $r=R(x^A)-\Delta r$ is trapped for small $\Delta r$.
We use equation (\ref{eq:gravity}) to generalize the definition of
trapping gravity to an arbitrary slice of a inwardly converging null
hypersurface. Then slices of positive trapping gravity tend toward
trapping as $r$ decreases.  However, in general, there seems to be no
purely local criterion which guarantees the existence of a trapped
surface before encountering a caustic as $r\rightarrow 0$.

In the special case of a spherically symmetric slice of a spherically
symmetric null cone, the Laplace equation for a marginally trapped
slice reduces to the algebraic condition that $V=0$, which is satisfied
where the $r=const$ hypersurface becomes null. The vacuum Schwarzschild
metric in ingoing null coordinates (which are equivalent to ingoing
Eddington-Finkelstein (IEF) coordinates) is given by $\beta=0$,
$V=-(r-2M)$, $U^A=0$ and $h_{AB}=q_{AB}$. In this case, $V=0$
determines the location of the event horizon (which coincides with the
apparent horizon) and the surface gravity reduces to
$\kappa=(4M)^{-1}$.  In the non-vacuum spherically symmetric case,
$V=0$ determines the apparent horizon.

In the absence of spherical symmetry, $\Theta_l$ vanishes on a slice of the
form $R=const$ at points for which $Q=0$, where
\begin{equation}
   Q =-V+{r^2\over \sqrt{q}}(\sqrt{q} U^A)_{,A}.
    \label{eq:qmarg} 
\end{equation}
We will refer to the largest $r=const$ slice of ${\cal N}_0$ on which
$Q\le 0$ as a ``Q-boundary'' ${\cal Q}$, relative to ${\cal S}_0$.
(${\cal S}_0$ enters here because it provides the reference for
$r=const$ slices).  Such a slice is everywhere trapped or marginally
trapped so that the Q-boundary provides a simple algebraic procedure
for locating an inner boundary inside an event horizon. A Q-boundary,
when it exists, will always lie inside (smaller $r$) or tangent to the
trapping horizon. Let $r=R_Q$ describe the Q-boundary and let $R_{min}$
($R_{max}$) be the smallest (largest) value of $r$ on the trapping
horizon. Then on the trapping horizon $R_{,A}=0$ at $R_{min}$ and
$R_{max}$ so that equation (\ref{eq:divergence}) implies $Q\ge 0$ at
$R_{min}$ (and $Q\le 0$ at $R_{max})$. Consequently, $R_Q \le R_{min}$,
with equality holding only when $D^2 R=0$ at $R_{min}$.

There are thus two possible strategies for positioning an inner boundary,
both of which ensure that the ignored portion of spacetime cannot causally
effect the exterior spacetime: (I) Use the trapping horizon, in which case
the 2D elliptic equation (\ref{eq:divergence}) must be solved on a sphere
in order to determine its location; or (II) use the Q-boundary which
is determined by a simple algebraic condition.

Strategy (I) is similar to approaches used to locate an apparent
horizon on a Cauchy hypersurface. The advantage in the null cone case
is that there is a natural radial coordinate defined by the coordinate
system to reduce the elliptic problem to 2 angular dimensions and to
define a mask for moving the excised region through the computational
grid. Strategy (II) carries no essential computational burden since the
quantities $V$ and $U$ are obtained by means of an inward radial
integral as part of the evolution scheme. One merely stops the
integration when the inequality defining the Q-boundary is satisfied.
Thus strategy (II) is preferable unless either caustics or
singularities appear before reaching the Q-boundary.  It is easy to
choose black hole initial data so that the Q-boundary and trapping
horizon agree at the beginning of the evolution but whether the
Q-boundary will move too far inward to be useful is a critical question
which would depend upon choices of lapse, shift and the geometry of the
matching world tube. Further research is necessary to decide if
strategy (II) is viable on geometric grounds in a highly asymmetric
spacetime.

\section{Collapse of a spherically symmetric scalar wave}
\label{sec:spherical}

\begin{figure}
\centerline{\epsfxsize=3.2in\epsfbox{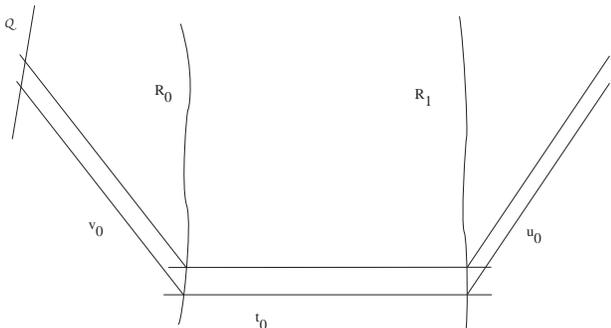}}
\caption{
This figure shows the two matching world tubes and the three coordinate patches.
}
\label{fig:spacetime}
\end{figure}

Our purpose here is to present a spherically symmetric model demonstrating the
feasibility of a stable global algorithm based upon three regions which
cover the spacetime exterior to a single black hole.
FIG \ref{fig:spacetime} illustrates 1-dimensional radial geometry.
The innermost region is evolved using an ingoing null algorithm whose
inner boundary $Q$ lies at the apparent horizon and whose outer boundary $R_0$
lies outside the black hole at the inner boundary of a region evolved
by a Cauchy algorithm.  Data is passed between these regions using a
matching procedure which is detailed below.  The outer boundary $R_1$ of the
Cauchy region is handled by matching to an outgoing null evolution.  The
details of the outgoing null algorithm \cite{scal1992} and of the
Cauchy evolution \cite{marsachoptuik1996} are not discussed since they have been
presented elsewhere. We will discuss the matching conditions
since they differ from those used previously \cite{gomez1996} due to different choices of gauge conditions. We will also present the field equations since they
are important for understanding the matching procedure.

The Cauchy evolution is carried out in IEF
coordinates.  The metric in this coordinate system is
\begin{equation}\label{eq:iefmet}
ds^2=\tilde{a}^2 \left( 2\tilde{\beta}-1 \right) d\tilde{t}^2 +
2\tilde{a}^2\tilde{\beta} d\tilde{t} d\tilde{r} + \tilde{a}^2 d\tilde{r}^2
+\tilde{r}^2 d\Omega^2.
\end{equation}
The set of equations used in the evolution are
\begin{equation}\label{eq:mom}
 K^{\theta}{}_{\theta}' + \frac{ K^{\theta}{}_{\theta} -
K^r{}_r}{\tilde{r}} -
\frac{4\pi\Phi\Pi}{\tilde{a}}=0 ,
\end{equation}
\begin{equation}\label{eq:betev}
\tilde{\beta} = \frac{\tilde{r}\tilde{a}K^{\theta}{}_{\theta}}{1 +
\tilde{r}\tilde{a}K^{\theta}{}_{\theta}} ,
\end{equation}
\begin{equation}\label{eq:aev}
\dot{\tilde{a}}=-\tilde{a}^2\left( 1-\tilde{\beta}\right) K^r{}_r +\left(
\tilde{a}\tilde{\beta}\right)' ,
\end{equation}
\begin{eqnarray}\label{eq:kttev}
\dot{ K^{\theta}{}_{\theta}}&=&\tilde{\beta} K^{\theta}{}_{\theta}' +
\tilde{a}\left(
1-\tilde{\beta}\right) K^{\theta}{}_{\theta}\left( K^r{}_r+2
K^{\theta}{}_{\theta}\right)
\nonumber \\ && +\frac{1-\tilde{\beta}}{\tilde{r}^2}\left(
\tilde{a}-\frac{1}{\tilde{a}}\right)+\frac{\tilde{\beta}'}
{\tilde{a}\tilde{r}} ,
\end{eqnarray}
\begin{equation}\label{eq:phiev}
\dot{\Phi}=\left(\tilde{\beta}\Phi+\left( 1-\tilde{\beta}\right)\Pi\right)' ,
\end{equation}
\begin{equation}\label{eq:piev}
\dot{\Pi}=\frac{1}{\tilde{r}^2}\left[
\tilde{r}^2\left(\tilde{\beta}\Pi+\left(
1-\tilde{\beta}\right)\Phi\right)\right]',
\end{equation}
where the over-dot represents partial with respect to $\tilde{t}$, prime
denotes
partial with respect to $\tilde{r}$ and the scalar field variables are
defined by
\begin{equation}\label{eq:defPhi}
\Phi \equiv \phi' \qquad
\Pi \equiv \frac{1}{1-\tilde{\beta}}\left(\dot{\phi} - \tilde{\beta} \phi
'\right).
\end{equation}

The outgoing-null metric is
\begin{equation}\label{eq:sphoutmet}
ds^2=-e^{2\beta} {V\over r} du^2 - 2 e^{2\beta} du dr + r^2 d\Omega^2,
\end{equation}
The outgoing hypersurface equations (\ref{eq:beta}) - (\ref{eq:v}) reduce to
\begin{eqnarray}
\beta_{,r} &=& 2\pi r \phi_{,r}^2,
\label{eq:sbeta}
\\
V_{,r} &=& e^{2\beta},
\label{eq:sV}
\end{eqnarray}
with $U^A=0$ and $h^{AB}=q_{AB}$,
and the outgoing version of the  scalar wave equation is
\begin{equation}
2(r\phi)_{,ru} = {1\over r}(rV\phi_{,r})_{,r}.
\label{eq:oscfield}
\end{equation}

An ingoing null evolution algorithm can be obtained from the outgoing
algorithm by the procedure described in Sec. \ref{sec:null}. Given the
ingoing null metric
\begin{equation}\label{eq:sphinmet}
ds^2=e^{2\beta} {V\over r} dv^2 + 2 e^{2\beta} dv dr + r^2 d\Omega^2,
\end{equation}
an independent set of
equations are the hypersurface equations
\begin{eqnarray}
\beta_{,r} &=& 2\pi r \phi_{,r}^2,
\label{eq:sibeta}
\\
V_{,r} &=& -e^{2\beta},
\label{eq:siv}
\end{eqnarray}
and the scalar wave equation
\begin{equation}
   2(r\phi)_{,rv}={1\over r}(rV\phi_{,r})_{,r}.
\label{eq:SWE}
\end{equation}
Given data for $\phi$ on ${\cal N}_0$ and on the world tube ${\cal R}_0$,
defined by $r=R_0$, and integration constants for $\beta$ and $V$ on ${\cal
R}_0$ evolution proceeds to null hypersurfaces ${\cal N}_v$, defined by
$v=const$, by an inward radial march along the null rays emanating inward
from ${\cal R}_0$.

\subsection{Initial data} \label{sec:idata}

The initial data consists of a Schwarzschild black hole of mass $M$ which
is
well separated from a Gaussian pulse of (mostly) ingoing scalar radiation.
Initially there is no scalar field present on the ingoing-null patch, so
the
initial data there is simply
\begin{eqnarray}
\phi &=& 0 ,\\
\beta &=& 0 ,\\
V &=& 2M-r.
\end{eqnarray}

Similarly, initially there is no scalar field on the outgoing-null patch,
so
we have $\phi=0$. The values for $\beta$ and $V$ are determined by matching
to the Cauchy data at the world tube (see below) and integrating the
hypersurface equations (\ref{eq:sibeta}) and (\ref{eq:siv}).

In the Cauchy region, the scalar field is given by
\begin{eqnarray}
\phi &=& A \tilde{r}
\exp{\left[ -\left(\tilde{r}-c\right)^d / \sigma^d \right] }, \\
\Phi &=& \phi \left[ {1\over\tilde{r}} -
\frac{d\left(\tilde{r}-c\right)^{d-1}}{\sigma^d}\right], \\
\Pi &=& \phi \left[ \frac{2-\tilde{\beta}}
{\tilde{r}\left( 1-\tilde{\beta}\right)}
- \frac{d\left(\tilde{r}-c\right)^{d-1}}{\sigma^d} \right],
\end{eqnarray}
where $A$, $c$, $d$, and $\sigma$ are scalars representing the pulse's
amplitude, center, shape, and width, respectively.  The geometric variables
are initialized using an iterative procedure, as detailed in
\cite{marsachoptuik1996}.

\subsection{Matching conditions} \label{sec:matchcond}

Since the IEF coordinate system is based on ingoing null cones, it is
possible to construct a simple coordinate transformation which maps the IEF
Cauchy metric to the ingoing null metric, namely
\begin{equation} \label{eq:incoordtr}
\tilde{t}=v-r, \qquad \tilde{r}=r.
\end{equation}
This results in the following transformations between the two metrics.
\begin{eqnarray} \label{eq:betain}
\beta &=& 1/2 \log{\left[ \tilde{a}^2\left( 1- \tilde{\beta} \right)
\right] } \\
\label{eq:Vin}
V &=& r \frac{2\tilde{\beta} - 1}{ 1- \tilde{\beta} } \\
\label{eq:tbetain}
\tilde{\beta} &=& \frac{V+r}{V+2r} \\
\label{eq:tain}
\tilde{a} &=& e^\beta \sqrt{V/r + 2}.
\end{eqnarray}

The extrinsic curvature components can be found using only the Cauchy
metric,
\begin{eqnarray}
K^{\theta}{}_{\theta} &=& \frac{\tilde{\beta}}{\tilde{r}\tilde{a}\left(
1-\tilde{\beta}\right)} \label{eq:Kttdef} \\
K^r{}_r &=& \frac{1}{\tilde{a}\left( 1-\tilde{\beta}\right)}\left(
\tilde{\beta}' +
\tilde{\beta}{ \tilde{a}' \over \tilde{a}} - { \dot{\tilde{a}} \over
\tilde{a}}
\right) .\label{eq:Krrdef}
\end{eqnarray}

Note that these transformation equations are valid everywhere in the spacetime,
not just at the world tube.

The matching conditions at the outer world tube are more complicated.
Both the Cauchy and characteristic systems share the same surface area
coordinate $r$ but there is no universal transformation between their
corresponding time coordinates. However, we can construct a coordinate
transformation which is valid everywhere on the world tube.  To do
this, we start with a general, differential coordinate transformation,
whose unknown function $F$ is to be determined from the matching
conditions:
\begin{equation}\label{eq:gentr}
d\bar{u} = F(\tilde{t}) d\tilde{t} - d\tilde{r}, \qquad dr=d\tilde{r}.
\end{equation}

To keep the notation simpler, we will write the Cauchy metric as
\begin{equation}
ds^2=g_{00}d\tilde{t}^2 + 2g_{01} d\tilde{t}d\tilde{r} + g_{11}
d\tilde{r}^2 +
\tilde{r}^2 d\Omega^2.
\end{equation}
Inverting (\ref{eq:gentr}) and substituting, we get
\begin{eqnarray}
ds^2&=&g_{00}\frac{1}{F^2} d\bar{u}^2 + 2\left( \frac{1}{F} g_{01} +
\frac{1}{F^2}
g_{00}\right) d\bar{u} dr 
\nonumber \\ &&+ \left( g_{11} + \frac{2}{F} g_{01} + \frac{1}{F^2}
g_{00} \right) dr^2 + r^2 d\Omega^2.
\end{eqnarray}

On the world tube, we require $u=\tilde{t}, so that \quad
du=d\tilde{t}$.  Thus, we set $du=d\bar{u}/F$.  For the metric, we get
\begin{eqnarray}
ds^2&=&g_{00} du^2 + 2\left( g_{01} + \frac{1}{F} g_{00}\right) du dr 
\nonumber \\ &&+ \left( g_{11}
+ \frac{2}{F} g_{01} + \frac{1}{F^2} g_{00} \right) dr^2 + r^2 d\Omega^2.
\end{eqnarray}

The condition that the $r$-direction be null implies that
\begin{equation}
g_{11} + \frac{2}{F} g_{01} + \frac{1}{F^2} g_{00} = 0.
\end{equation}
Upon substitution of the IEF metric functions, this determines that
\begin{equation}
F = 1 - 2\tilde{\beta}.
\end{equation}

The matching conditions are then
\begin{eqnarray} \label{eq:betaout}
\beta &=& 1/2 \log{\left[ \tilde{a}^2\left( 1- \tilde{\beta} \right)\right]
} \\
\label{eq:Vout}
V &=& r \frac{1 - 2\tilde{\beta}}{ 1- \tilde{\beta} } \\
\label{eq:tbetaout}
\tilde{\beta} &=& \frac{V-r}{V-2r} \\
\label{eq:taout}
\tilde{a} &=& e^\beta \sqrt{2 - V/r}.
\end{eqnarray}

\subsection{Finite difference implementation}

As is typical with finite difference calculations, the continuum functions
are discretized spatially and placed on grids with $N$ points.  In our case
we have three regions.  The inner-null variables are placed on grids with
$N_i$ points, the Cauchy variables on grids with $N_c$ points, and the
outer-null variables on grids with $N_o$ points.  On each grid, the spatial
index $i$ runs from $1$ to $N_i$, $N_c$ or $N_o$, respectively, with $1$ representing the smallest $r$ value.

\begin{figure}
\centerline{\epsfxsize=3.2in\epsfbox{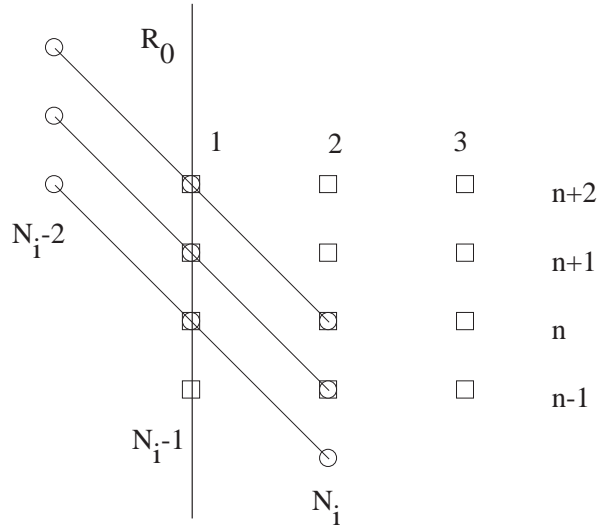}}
\caption{
This diagram shows the finite difference grids around the inner world
tube. The squares represent the Cauchy grid, while the circles are the
null grid.  The diagonal lines are the ingoing null cones, and the
vertical line $R_0$ is the inner world tube.  The innermost points of
the Cauchy grid lie on the world tube, while the null grid extends
outside the world tube.  Notice that the grids align in space and
time.
}
\label{fig:inmatch}
\end{figure}

In addition to the spatial discretization, each function needs two or
more time levels.  While both the Cauchy and null evolution schemes use
only two time levels, we keep an extra level in each to facilitate the
matching.  FIG \ref{fig:inmatch} shows how the finite difference grids
match at the inner world tube.  The two grids are aligned in both $r$
and $\tilde{t}$.  This means no interpolations are necessary.  The
world tube is at $i=1$ on the Cauchy grid and $i=N_i-1$ on the ingoing
null grid.  The Cauchy variables need boundary values on time level
$n+1$ at $i=1$.  The metric values come from the null variables at
level $n+1$, with $i=N_i-1$, using the transformation equations
(\ref{eq:tbetain}) and (\ref{eq:tain}).  These relations are algebraic
and straightforward to implement.  Boundary values for the extrinsic
curvature components come from equations (\ref{eq:Kttdef}) and
(\ref{eq:Krrdef}).  $K^\theta{}_\theta$ is computed algebraically and
$K^r{}_r$ is computed using second-order, centered-in-time,
forward-in-space derivatives in the Cauchy grid.

The transformation of the scalar field requires transforming derivatives
between the two coordinate patches.  At the world tube, the relationships
among the derivatives are
\begin{eqnarray}
\partial_r &=& \partial_{\tilde{r}} - \partial_t \\
\partial_{\tilde{r}} &=& \partial_r + \partial_v \\
\partial_v &=& \partial_t .
\end{eqnarray}
These lead to the following equations for $\Phi$ and $\Pi$:
\begin{equation}
\Phi^{n+1}_1 = \frac{1}{\tilde{r}_1} \frac{g^{n+2}_{N_i -1} - g^{n}_{N_i
-1}}{2 dv} + \frac{1}{\tilde{r}_1} \frac{g^{n+1}_{N_i} - g^{n+1}_{N_i
-2}}{2 dr} -
\frac{g^{n+1}_{N_i -1}}{\tilde{r}_1 ^2}
\end{equation}
and
\begin{equation}
\Pi^{n+1}_1 = \frac{1}{1-\tilde{\beta}^{n+1}_1} \left(
\frac{1}{\tilde{r}_1} \frac{g^{n+2}_{N_i -1} - g^{n}_{N_i -1}}{2 dv} -
\tilde{\beta}^{n+1}_1 \Phi^{n+1}_1 \right).
\end{equation}

The null variables need boundary values at $n+2$, $i=N_i$.  The metric
values come from the Cauchy variables at level $n$, with $i=2$, using
(\ref{eq:betain}) and (\ref{eq:Vin}). The evolution equation for $g$ at
the world tube is
\begin{eqnarray}
g^{n+2}_{N_i} &=& g^{n+2}_{N_i} + r_{N_i} dv \left[ \left( 1 -
\frac{\tilde{\beta}^n_2 + \tilde{\beta}^{n-1}_2}{2} \right) \frac{\Pi^{n}_2
+ \Pi^{n-1}_2}{2} \right.
\nonumber \\ && \left. + \frac{\tilde{\beta}^n_2 + \tilde{\beta}^{n-1}_2}{2}
\frac{\Phi^{n}_2 + \Phi^{n-1}_2}{2}\right].
\end{eqnarray}

\begin{figure}
\centerline{\epsfxsize=3.2in\epsfbox{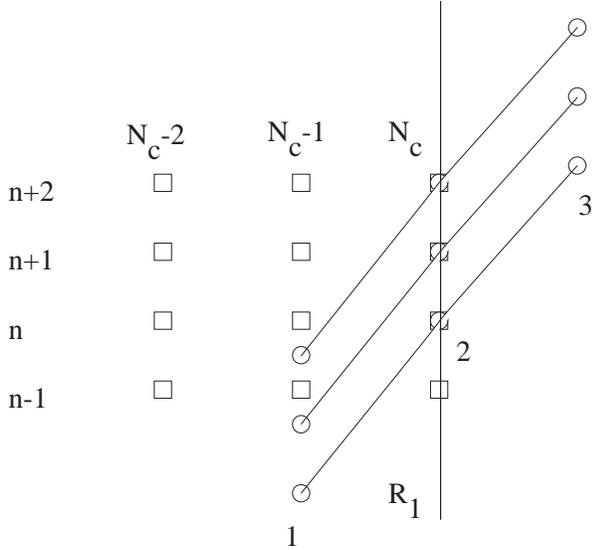}}
\caption{
This diagram shows the finite difference grids around the outer world
tube. The squares represent the Cauchy grid, while the circles are the
null grid.  The diagonal curves are the outgoing null cones, and the
vertical line $R_1$ is the outer world tube.  The outermost points of
the Cauchy grid lie on the world tube, while the null grid extends
inside the world tube.  Notice that the the grids align in time only on
the world tube, but in space at the world tube and just inside.
}
\label{fig:outmatch}
\end{figure}

The situation at the outer world tube is shown in FIG \ref{fig:outmatch}.
Here, the grids align in space at two values of $r$ but in time only at the
world tube.  The Cauchy metric boundary values at $i=N_c$ come directly
from the null variables at $i=2$ using the transformation equations
(\ref{eq:tbetaout}) and (\ref{eq:taout}).  Since the grids don't align in
time as they do at the inner world tube, we use a different procedure for
the scalar field boundary values.  We use the derivative transformation
\begin{equation}
\partial_r = \frac{1}{1-2\tilde{\beta}} \partial_t + \partial_{\tilde{r}}
\end{equation}
to get an evolution equation for $\phi$ at the world tube.  We then set
$\Phi$ and $\Pi$ using their definitions (\ref{eq:defPhi}) and backward 
second-order derivatives.

The boundary values for the null variables must be interpolated in time
using the Cauchy variables at $n$ and $n-1$. Given values for a
function $f$ at time levels $n$ and $n-1$, we can get its value $f^s$
at the beginning of the null cone using
\begin{eqnarray}
f^s &=& f^n \left( 3 - \frac{dr}{dt} \frac{1}{1-2\tilde{\beta}^{n+2}_{N_c-1}}
\right) 
\nonumber \\ &&+ f^{n-1} \left( -2 + \frac{dr}{dt}
\frac{1}{1-2\tilde{\beta}^{n+2}_{N_c-1}} \right).
\end{eqnarray}
Notice that this expression requires a value for $\tilde{\beta}$ at level
$n+2$, something that won't be known until the next time step.  We have
found it sufficient to extrapolate from the previous time levels using
\begin{equation}
\tilde{\beta}^{n+2}_{N_c-1} = 3 \tilde{\beta}^{n+1}_{N_c-1} - 3
\tilde{\beta}^{n}_{N_c-1} + \tilde{\beta}^{n-1}_{N_c-1}.
\end{equation}

Thus, to get values for $\beta$ and $V$ we use the matching conditions
(\ref{eq:betaout}) and (\ref{eq:Vout}) along with the above
interpolation.  For the scalar field, we interpolate the value of $g$
from the Cauchy grid, using $g=r\phi$.

The apparent horizon is found on the ingoing null cones using the apparent
horizon equation which reduces simply to $V=0$.  When the scalar field
passes into the black hole, the horizon grows outward and we simply stop
evolving the grid points that are now inside.

\subsection{Performance}\label{sec:compare}

To evaluate the performance of this approach, we compare it to a second
order accurate, purely Cauchy evolution in IEF coordinates, as
presented in \cite{marsachoptuik1996}.  For the comparison shown here,
we place the outer boundary of the Cauchy evolution at $r=62M$ and
evolve to $t=40M$ to prevent any outer boundary effects from
influencing the comparison.  The scalar field pulse is centered at
$r=22M$ and has a mass of $0.5M$.

\begin{figure}
\centerline{\epsfxsize=3.2in\epsfbox{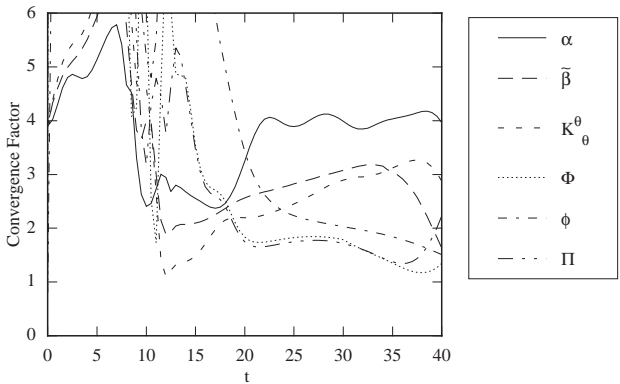}}
\caption{
This plot shows the mutual convergence of the Cauchy variables from the
matching evolution and the pure Cauchy evolution.  The evolutions are
for a strong, ingoing scalar pulse ($0.5M$) outside a black hole.  A
convergence factor near four indicates second order convergence, while
one near two indicates first order convergence.
}
\label{fig:converg}
\end{figure}

FIG \ref{fig:converg} shows the mutual convergence between the Cauchy
variables in the two codes.  This demonstrates that the two programs
are solving the same problem, and provides evidence that the matching
approach generates the correct spacetime.

The reason that the convergence rate appears to drop to first order
when the scalar field hits the horizon is an artifact arising from the
motion of the horizon. As mass falls into the hole, there is a critical
amount which causes the horizon to move out by one grid point.  If at
the coarsest resolution the horizon moves a distance $dr$ then on the
next finer grid it only moves by $dr/2$, and so on.  Thus, at different
resolutions, the black holes have slightly different locations. The
resulting shift in the location of the inner boundary causes
convergence between successively finer numerical solutions to drop from
second order to first order. This is an unavoidable diagnostic effect
due to the comparison of numerical solutions. We believe that the
numerical solution would converge at second order to an exact solution
of the physical problem. No exact solutions are known to use for a
check of this, but for a weak scalar field the horizon does not move and
we do measure second order convergence throughout the evolution.  Further,
we see the same convergence order drop for strong fields in Cauchy-only or
null-only evolutions, and thus are certain it is not due to the matching
procedure.

\section{Conclusions}

Our work shows that the matching approach provides as good a solution
to the black hole excision problem in spherical symmetry as 
previous treatments \cite{scheel1995a,scheel1995b,marsachoptuik1996,anninos1995}. 
It also has
some advantages over the pure Cauchy approach, namely, it is
computationally more efficient (fewer variables), and is much easier to
implement.  We achieved a stable evolution simply by transforming the
outgoing-null evolution scheme to work on ingoing null cones, and
implementing it.  Achieving stability with a purely Cauchy scheme in
the region of the apparent horizon is trickier, involving much trial
and error in choosing difference schemes.  It should be noted, however,
that implementing the matching may be tricky, especially in higher
dimensions.  Whether it is easier than implementing Cauchy differencing
near the horizon remains to be seen.

Also, using the ingoing null formulation, we have achieved the stable
evolution of a Schwarzschild black hole in 3-dimensions (the details
will be presented elsewhere) and are working on rotating and moving
black holes.  Long term stable evolution of a 3D black hole has yet to
be demonstrated with a Cauchy evolution.

\begin{center}
{\bf ACKNOWLEDGMENTS}
\end{center}

\vspace{0.3cm}
This work has been supported by NSF PHY 9510895 to the University of
Pittsburgh and by the Binary Black Hole Grand Challenge Alliance, NSF
PHY/ASC 9318152 (ARPA supplemented). Computer time for this project has
been provided by the Pittsburgh Supercomputing Center under grant
PHY860023P. We thank R. A. Isaacson for helpful comments on the
manuscript.

\end{document}